\documentclass[iop,apj,revtex4]{emulateapj}

\usepackage[dvips]{color}
\usepackage{graphicx}
\usepackage{mathrsfs}
\usepackage{color}

\begin{document}

\title{The Composition of Interstellar Grains Toward $\zeta$ Ophiuchi: Constraining the Elemental Budget Near the Diffuse-Dense Cloud Transition}

\author{Charles A.\ Poteet$^{1}$, Douglas C.\ B.\ Whittet$^{1}$, and Bruce T.\ Draine$^{2}$}

\affil{$^{1}$New York Center for Astrobiology, Department of Physics, Applied Physics and Astronomy,\\ Rensselaer Polytechnic Institute, 110 Eighth Street, Troy, NY 12180, USA; charles.poteet@gmail.com\\
$^{2}$Princeton University Observatory, Peyton Hall, Princeton, NJ 08544, USA} 

\slugcomment{To be published in the Astrophysical Journal: received 2014 August 18; accepted 2015 January 12}

\begin{abstract}

We investigate the composition of interstellar grains along the line of sight toward \object{$\zeta$ Ophiuchi}, a well-studied environment near the diffuse-dense cloud transition.   A spectral decomposition analysis of the solid-state absorbers is performed using archival spectroscopic observations from the \emph{Spitzer Space Telescope} and \emph{Infrared Space Observatory}.  We find strong evidence for the presence of sub-micron-sized amorphous silicate grains, principally comprised of olivine-like composition, with no convincing evidence of H$_{2}$O ice mantles.  However, tentative evidence for thick H$_{2}$O ice mantles on large ($a \approx$ 2.8~$\micron$) grains is presented.  Solid-state abundances of elemental Mg, Si, Fe, and O are inferred from our analysis and compared to standard reference abundances.  We find that nearly all of elemental Mg and Si along the line of sight are present in amorphous silicate grains, while a substantial fraction of elemental Fe resides in compounds other than silicates.  Moreover, we find that the total abundance of elemental O is largely inconsistent with the adopted reference abundances, indicating that as much as $\sim$156~ppm of interstellar O is missing along the line of sight.  After taking into account additional limits on the abundance of elemental O in other O-bearing solids, we conclude that any missing reservoir of elemental O must reside on large grains that are nearly opaque to infrared radiation.          

\end{abstract}


\section{Introduction}\label{sec:intro}

The interchange of gas and dust plays an essential role in the evolution of the interstellar medium (ISM).  The surfaces of the dust grains act as both repositories for atoms and molecules adsorbed from the gas-phase and active sites for molecule formation over a wide range of interstellar environments.  The apparent under-abundances or ``depletions'' of many heavy elements detected in the gas-phase are interpreted as evidence for their inclusion on or in interstellar grains.  Forty years ago, \citet{Greenberg74} concluded that for the cosmically abundant elements of carbon, nitrogen, and oxygen, significantly larger amounts were missing from the gas-phase than could be accounted for by accretion onto interstellar dust grains.  In view of this critical finding, \citet{Greenberg74} hypothesized that {\it ``there is more hidden mass than there is mass in the dust''}, with such mass possibly existing in the form of intermediate-sized ``snowballs'' or more complicated interstellar molecules that had yet to be detected in the solid-state. 

With major advancements in the sensitivity and resolution of ultraviolet spectroscopy, as well as decreases in solar and stellar abundance estimates, apparent inconsistencies between the expected (approximately solar) and observed (gas and dust) interstellar abundances have largely subsided over the years \citep[e.g.,][]{Cardelli94, Sofia94, Snow96, Sofia01}.  However, the depletion of elemental O remains problematic in the more heavily depleted regions of the ISM.  In a recent comprehensive study, \citet{Jenkins09} presents a unified representation of depletions, in which previous results from the literature are re-analyzed using a consistent system of oscillator strengths for the various transitions used to derive gas-phase column densities.  Independent of the adopted reference abundance, \citet{Jenkins09} finds that elemental O (as observed in its dominant neutral atomic form, \ion{O}{1}, in the diffuse ISM) is depleted from the gas-phase at a rate that far exceeds the rate at which it can be sequestered into silicate and metallic oxide particles.

As later explored in detail by \citet{Whittet10}, this result implies a crisis in our understanding of interstellar dust.  At the high end of the density range covered by the depletion measurements (i.e., near the transition between diffuse neutral and dense molecular phases of the ISM), as much as 25\%--30\% of the total elemental O budget is unaccounted for in any known observed form.  Other than silicates and oxides, possible O-bearing reservoirs in common grain materials include organic refractory matter \citep{Li97} and interstellar ices of H$_{2}$O, CO, and CO$_{2}$.  However, ice mantles have only been detected in the denser regions of the ISM \citep[][and references therein]{Whittet10}, and a failure to detect the $\sim$6~$\micron$ C$\dbond$O carbonyl feature in diffuse ISM dust appears to place strong constraints on the abundance of O-bearing organic matter \citep{Pendleton02, Chiar13}.

\begin{figure}[htp]
\centering
\includegraphics*[width=0.482\textwidth]{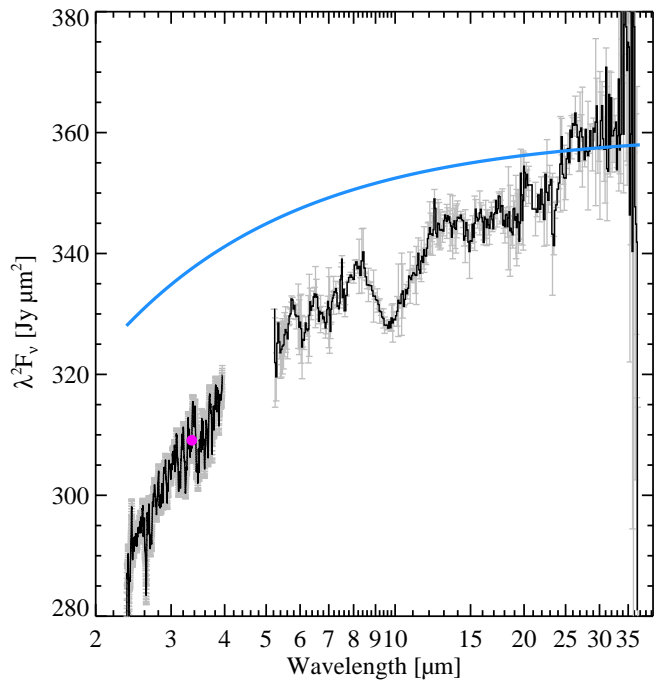}
\caption{Combined archival \emph{ISO}-SWS and \emph{Spitzer}-IRS spectra (black lines) of the $\zeta$ Oph line of sight and 3.4~$\micron$ broadband photometry (magenta circle) from \emph{WISE}, shown with their respective uncertainties (gray lines).  For comparison, the expected photospheric continuum emission by an O9.5 \textrm{V} star (blue line), corresponding to a $T = 34300$~K blackbody \citep{Howarth01}, is also indicated.  Note that in the Rayleigh-Jeans limit, a blackbody spectrum has $\lambda^{2}F_{\nu}$ equal to a constant.}
\label{fig:spectra}
\end{figure}

A further possibility is that O-bearing gaseous molecules, such as CO and O$_{2}$, might be important contributors.  However, in the case of CO, the observed abundances toward diffuse molecular clouds are well-constrained, and clearly too small to account for a significant fraction of the total elemental O abundance \citep{Liszt09}.  Abundances of O$_{2}$ are less well-constrained, but are unlikely to exceed those of CO.  Finally, the possible existence of large, micron-sized interstellar grains have also been proposed as viable O-bearing repositories \citep{Greenberg74, Jenkins09}, in view of the known difficulty in detecting such materials.   

In this paper, we re-examine the line of sight toward the bright O9.5 \textrm{V} star \object{$\zeta$ Ophiuchi} \citep[\object{$\zeta$ Oph}, \object{HD 149757};][]{Herbig68}. The low-velocity ($v_{\odot} = -15$~km s$^{-1}$) component toward this well-studied star is regarded as a prototypical ``cool diffuse" cloud with high elemental depletions and partial \ion{H}{1} $\rightarrow$ H$_2$ conversion, and is representative of an environment near the diffuse-dense ISM transition \citep{Savage96, Jenkins09, Liszt09}.  We utilize archival infrared spectra from the \emph{Spitzer Space Telescope} \citep{Werner04} and the \emph{Infrared Space Observatory} \citep[\emph{ISO};][]{Kessler96} to quantify the solid-state absorption by refractory and molecular compounds along the line of sight, and combine our results with previously published data from the literature to assess the overall elemental budget toward \object{$\zeta$ Oph}.

In general, $\sim$20--29\% of the elemental O abundance in the ISM is expected to be present in amorphous silicate grains.  The chemical and physical properties of amorphous silicates can be determined by detailed modeling of their observed 10 and 20~$\micron$ resonance features.  Revealed through infrared spectroscopy, it is well-known that the profile of the 10~$\micron$ silicate emission feature detected toward the Trapezium \citep{Forrest75} is similar to the silicate absorption feature observed toward dense molecular clouds \citep{Gillett75, Roche84}, whereas silicates in diffuse clouds possess a narrower absorption profile \citep{Roche84, Bowey98}.  However, the high signal-to-noise ratio of the \emph{Spitzer} observations toward \object{$\zeta$ Oph} allows us, for the first time, to characterize the composition of the silicates in a diffuse interstellar cloud near the \ion{H}{1} $\rightarrow$ H$_2$ transition.

In Section \ref{sec:obs}, an overview of the archival \emph{ISO} and \emph{Spitzer} observations is presented.  In Section \ref{sec:specdecomp}, spectral decomposition analyses of the infrared spectra toward \object{$\zeta$~Oph} are employed to investigate the presence of solid-state absorption along the line of sight.  In Section \ref{sec:abun}, we use our results to quantify the elemental abundances of Mg, Si, O, and Fe carried by amorphous silicate grains, and also place strict limits on the abundance of elemental O in other O-bearing solids.  Finally, implications of our results with regard to the ``missing oxygen" problem are discussed in Section \ref{sec:discuss}. 

\section{Archival Observations}\label{sec:obs}

Observations toward \object{$\zeta$ Oph} are assembled using archival spectra obtained by the \emph{Spitzer Space Telescope} Infrared Spectrograph \citep[IRS;][]{Houck04} and the Short Wavelength Spectrometer \citep[SWS;][]{de Graauw96} on board \emph{ISO}.  As part of the \emph{Spitzer} Atlas of Stellar Spectra \citep[PID 485;][]{Ardila10}, the \emph{Spitzer}-IRS observations (\emph{Spitzer} AOR key 27582976) were carried out with the low-resolution ($R$ $=$ 60--120) modules, Short-Low (SL; 5.2--14.0~$\micron$) and Long-Low (LL; 14.0-36.1~$\micron$), on 2008 September 10.  The observations utilized the standard two-position nod configuration, yielding a total on-source exposure time of 12 s for each module.  The extracted 5--36~$\micron$ \emph{Spitzer}-IRS spectrum was retrieved from the Cornell Atlas of \emph{Spitzer} IRS Sources \citep[CASSIS v5.2;][]{Lebouteiller11}, which is based on the Advanced Optimal Extraction of the \emph{Spitzer} Science Center S18.18.0 pipeline basic calibrated data \citep{Lebouteiller10}.  Post-extraction processing was performed in order to scale the flux density of the SL module to that of the LL module at 14~$\micron$.  The uncertainties in the flux density are estimated to be half the difference between the two independent spectra from each nod position.

Covering the nominal spectral range of 2.4--45~$\micron$, the \emph{ISO}-SWS full grating scans were performed on 1997 August 5 (\emph{ISO} TDT 62803102) with the SWS01 Astronomical Observing Template at speed~1.  The observations result in a mean spectral resolving power of $R$ $\approx$~200--300 and a total exposure time of 1140~s.  The extracted \emph{ISO}-SWS spectrum was retrieved from the SWS Atlas of fully processed spectra \citep{Sloan03}.  Utilizing broadband photometry from the \emph{Wide-field Infrared Survey Explorer} \citep[\emph{WISE};][]{Wright10} All-Sky Data Release, the \emph{ISO}-SWS spectrum was scaled by a factor of 1.01 to match the flux density derived from the \emph{WISE} 3.4~$\micron$ passband.

\begin{figure*}[htp]
\centering
\includegraphics*[width=1.005\textwidth]{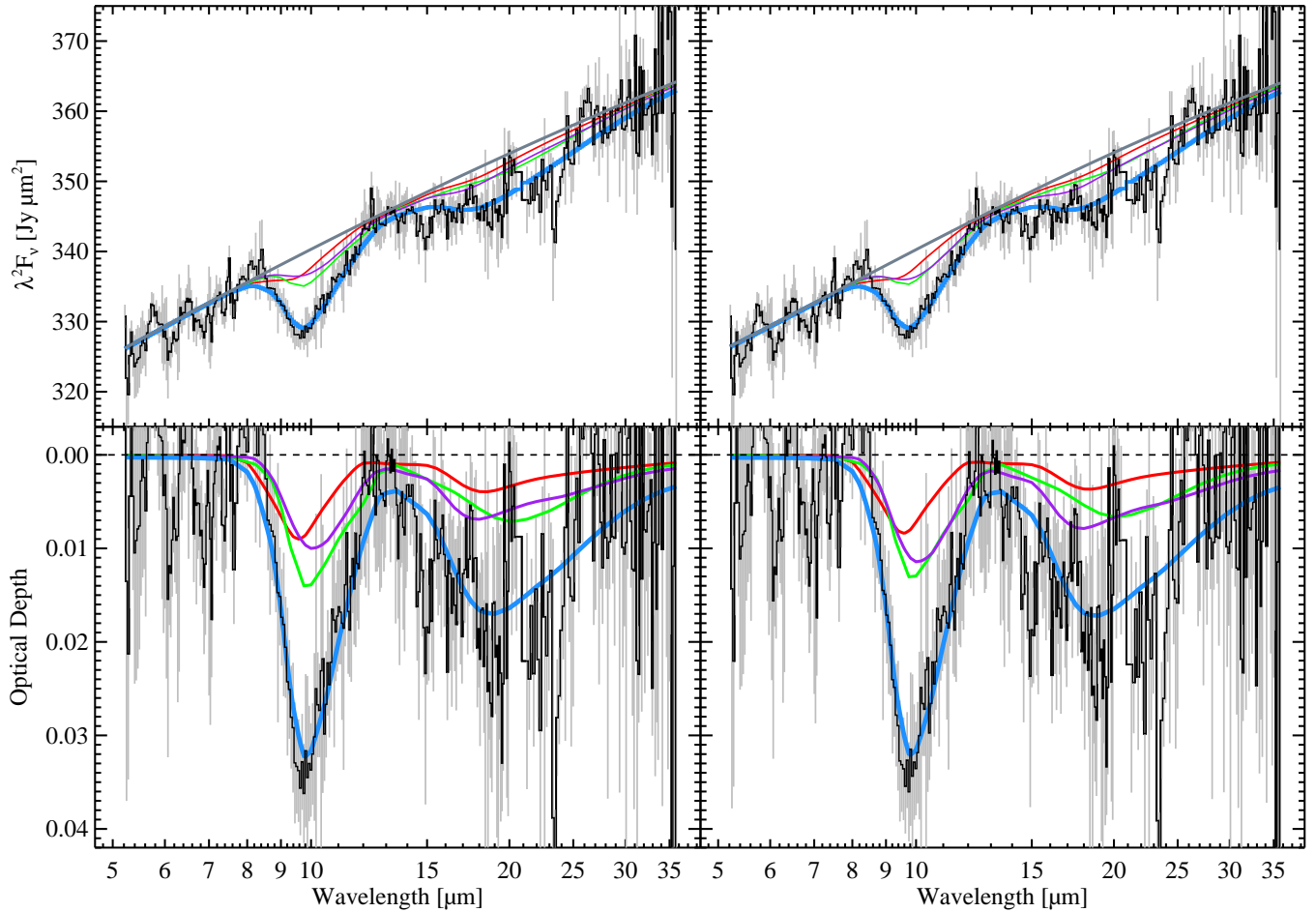}
\caption{\emph{Spitzer}-IRS spectrum and best-fit spectral decomposition models for the $\zeta$ Oph sightline.  Spectral decomposition models are shown for a second- (left panels) and fourth-order (right panels) polynomial flux continuum; the third-order spectral decomposition model results are identical to those of the fourth-order spectral decomposition model.  Top panels: observed spectrum (black line), model flux continuum (gray line), and best-fit model spectrum (blue line).  Bottom panels: observed optical depth spectrum (black line) and best-fit model optical depth (blue line).  The total absorption is modeled as a linear combination of amorphous silicates with Mg$_{2}$SiO$_{4}$ (green line), MgFeSiO$_{4}$ (violet line), and MgSiO$_{3}$ (red line) stoichiometries.}
\label{fig:models}
\end{figure*}

The combined 2--36~$\micron$ spectra are shown in Figure \ref{fig:spectra}.  However, due to the relatively poor signal-to-noise ratio in the \emph{ISO}-SWS spectrum at wavelengths longward of 4~$\micron$, only the 2.4--4~$\micron$ spectral range is retained to investigate the presence of H$_{2}$O ice along the line of sight.  For comparison, the expected photospheric continuum emission by an O9.5 \textrm{V} star, corresponding to a $T = 34300$~K blackbody \citep{Howarth01}, is also indicated.  Note that in the Rayleigh-Jeans limit, a blackbody spectrum has $\lambda^{2}F_{\nu}$ equal to a constant.  Differences between the observed spectrum and that of a pure stellar photosphere are attributed to extinction by intervening dust, consisting of amorphous silicates and possibly other dust species lacking spectral features in the mid-infrared wavelength range (e.g., amorphous carbon or metallic iron).

\section{Spectral Decomposition Analysis}\label{sec:specdecomp}

In order to characterize the interstellar absorbers along the line of sight toward \object{$\zeta$ Oph}, we approximate the attenuation by dust as an intermediate screen against a bright infrared continuum.  Historically, infrared absorption features have been studied by modeling optical depth spectra with laboratory-derived spectra of candidate materials.  However, in this classical approach, optical depth spectra are often inferred through the subjective process of local-continuum fitting.  In consequence, this process may artificially set the underlying optical depth to be zero over a pre-defined spectral region, which can strongly affect the subsequent modeling of the absorption profiles.  In light of this concern, we choose to carry out a different method in extracting information on the absorbers.

\setlength{\tabcolsep}{0.099 in}
\begin{deluxetable*}{cccccccccc}
\centering
\tablecolumns{10}

\tablecaption{Summary of Best-fit Optical Depth Parameters Toward $\zeta$ Oph}

\tablehead{
\colhead{Poly.\tablenotemark{a}} &
\colhead{$\tau_{\rm{sil}}$} &
\colhead{$\tau_{\rm{sil}}$} &
\colhead{$\tau_{\rm{sil}}$} &
\colhead{$\tau_{\rm{sil}}$} &
\colhead{$\tau_{\rm{H_{2}O}}$} &
\colhead{$\tau_{\rm{Fe_{3}O_{4}}}$\tablenotemark{b}} &
\colhead{$\tau_{\rm{POM}}$\tablenotemark{c}} &
\colhead{$\chi_{\nu}^{2}$} & 
\colhead{$\nu$}\vspace{0.02 in}\\
\colhead{Degree} &
\colhead{(total)} &
\colhead{(Mg$_{2}$SiO$_{4}$)} &
\colhead{(MgFeSiO$_{4}$)} &
\colhead{(MgSiO$_{3}$)} &
\colhead{} &
\colhead{} &
\colhead{} &
\colhead{} &
\colhead{}  \vspace{0.025 in}\\
\cline{1-10}\vspace{-0.06 in} \\
\multicolumn{10}{c}{\emph{Spitzer}-IRS spectral decomposition models}}
\startdata
2 \dotfill & 0.032 $\pm$ 0.009 & 0.014 $\pm$ 0.006 & 0.010 $\pm$ 0.006 & 0.009 $\pm$ 0.004 & \nodata & $\leq$0.006 & $\leq$0.002 & 1.21 & 327 \\
3 \dotfill & 0.031 $\pm$ 0.011 & 0.013 $\pm$ 0.007 & 0.011 $\pm$ 0.007 & 0.008 $\pm$ 0.004 & \nodata & $\leq$0.006 & $\leq$0.002 & 1.21 & 326 \\
4 \dotfill & 0.031 $\pm$ 0.011 & 0.013 $\pm$ 0.007 & 0.011 $\pm$ 0.007 & 0.008 $\pm$ 0.004 &\nodata & $\leq$0.006 & $\leq$0.002 & 1.21 & 325\vspace{-0.03 in}\\
\cutinhead{\emph{ISO}-SWS spectral decomposition models}
2 \dotfill & \nodata & \nodata & \nodata & \nodata & $\leq$0.008\tablenotemark{d} & \nodata & \nodata & \nodata & \nodata \\
2 \dotfill & \nodata & \nodata & \nodata & \nodata & $\leq$0.009\tablenotemark{e} & \nodata & \nodata & 2.34 & 1957\vspace{0.015 in}
\enddata
\tablecomments{Uncertainties are based on statistical errors in the \emph{Spitzer}-IRS and \emph{ISO}-SWS spectra only.}
\tablenotetext{a}{Degree of the polynomial flux continuum.}
\tablenotetext{b}{Optical depth upper limit estimate of the 16~$\micron$ Fe$_{3}$O$_{4}$ feature.}
\tablenotetext{c}{Optical depth upper limit estimate of the 8.9~$\micron$ POM feature.}
\tablenotetext{d}{Optical depth upper limit estimate of the 3.1~$\micron$ H$_{2}$O ice feature for sub-micron-sized mantled grains.}
\tablenotetext{e}{Derived from the 95\% upper confidence limit of the 3.53~$\micron$ H$_{2}$O ice feature for thick mantles with central radius of $a =$ 2.8~$\micron$.}
\label{tab:parameters}
\end{deluxetable*}

In our approach, which is based on simulating the emergent spectral energy distribution (SED), the flux continuum and extinction opacity are simultaneously determined in the logarithmic domain by 
\begin{equation}\label{eq:sed}
\lambda F_{\lambda}^{\rm{obs}} = \lambda F_{\lambda}^{\rm{cont}} e^{-\sum_{i} \alpha_{i} \tau_{\lambda,i}},
\end{equation}
\noindent where $F_{\lambda}^{\rm{obs}}$ and $F_{\lambda}^{\rm{cont}}$ are the observed and continuum flux densities, respectively, $\tau_{\lambda,i}$ is the normalized optical depth of an individual absorbing component, and $\alpha_{i}$ is its corresponding non-negative scaling parameter.  The flux continuum is approximated as a low-order polynomial, $\log(\lambda F_{\lambda}^{\rm{cont}}) = \sum_{i} a_{i}(\log\lambda)^{i}$, with a maximum degree ranging from two to four, and represents the stellar photospheric contribution by \object{$\zeta$ Oph}, as well as any unknown featureless dust extinction.  The extinction optical depth is modeled using a linear combination of amorphous silicates and H$_{2}$O ice absorption spectra.  We assume that these individual absorbing components have fixed profiles, but their relative contribution to the total optical depth may vary.

We adopt amorphous silicate optical constants with stoichiometric compositions from the melting and quenching experimental studies of \citet{Dorschner95} and \citet{Henning96}.  In particular, we use Fe-free and Fe-bearing amorphous silicates with pyroxene ({MgSiO$_{3}$ and MgFeSi$_{2}$O$_{6}$) and olivine (Mg$_{2}$SiO$_{4}$ and MgFeSiO$_{4}$) compositions.  Following the statistical approach of \cite{Min03}, the optical properties of the amorphous silicate species are modeled using a distribution of hollow spheres (DHS) in the Rayleigh limit ($\lambda$ $\gg$ 2$\pi$$a$, where $a$ is the radius of the largest grains).  In DHS theory, the optical properties of hollow spherical particles are uniformly averaged from 0 $<$ $f$ $<$ $f_{\rm{max}}$, where $f$ is the fraction of the total volume occupied by the central vacuum inclusion.  The value of $f_{\rm{max}}$ indicates the degree of irregularity of the particles, with small values of $f_{\rm{max}}$ representing nearly spherical particles.  Adopting the maximum volume fraction of $f_{\rm{max}}$ = 0.7 determined for the \object{Sgr A$^{\ast}$} line of sight \citep{Min07}, the resulting ensemble of particles is representative of small, irregularly-shaped amorphous silicate grains.

For the H$_{2}$O ice component, we adopt optical constants derived from the laboratory spectrum of pure, amorphous H$_{2}$O ice \citep[$T = 10$ K;][]{Hudgins93}.  Because irregularly shaped particle models have been largely successful at simulating the profile shape of interstellar ice features \citep[e.g.,][]{Pontoppidan03}, the optical properties of H$_{2}$O ice are modeled using a continuous distribution of ellipsoids (CDE) in the Rayleigh limit \citep{Bohren83}.  However, we note that the calculated absorption cross section of H$_{2}$O ice is not very sensitive to the choice of the adopted particle shape model (i.e., only small differences in the profile shapes of the absorption features are found if we instead adopt small, homogenous spherical mantles).   

Utilizing the model described in Equation (\ref{eq:sed}) and the aforementioned solid-state absorption spectra, we employ a nonlinear least-squares fitting technique \citep{Markwardt09} to simulate the mid-infrared SED of \object{$\zeta$ Oph}.  When present, blended silicate and H$_{2}$O ice absorption features occupy nearly all of the mid-infrared spectral region, and for this reason the \emph{Spitzer}-IRS spectrum is modeled over the entire 5.3--34~$\micron$ wavelength range.  In the fitting procedure, the initial values of the polynomial coefficients of the flux continuum and the scaling parameter of each absorbing component are randomized, with parameter convergence typically occurring after 30 or fewer iterations.  Adopting a detection limit criterion of $\geq$1$\sigma$, the best-fit model parameters are evaluated for significance upon convergence.  If absorbing components are detected with a significance less than 1$\sigma$, their corresponding scaling parameters are set to zero and the fitting procedure is repeated until all remaining components satisfy the detection criterion.  The formal uncertainties reported by the fitting procedure are derived from the statistical uncertainties within the \emph{Spitzer}-IRS spectrum.  If the resultant model is of good quality (i.e., the chi-squared per degree of freedom, $\chi^{2}_{\nu}$, is of order unity), the estimated parameter uncertainties are subsequently scaled by the square root of $\chi^{2}_{\nu}$.  

The best-fit spectral decomposition model results toward \object{$\zeta$ Oph} are presented in Figure \ref{fig:models} for a second- and fourth-order polynomial flux continuum.  The \emph{Spitzer}-IRS spectrum and best-fit model spectrum are shown in the top panels, while the contributions by sightline absorbers to the observed optical depth spectrum are presented in the bottom panels.  The best-fit optical depth parameters are listed in Table \ref{tab:parameters} for models using second-, third-, and fourth-order polynomial flux continua.  

\subsection{Amorphous Silicate Composition}\label{sec:silcomp}

In general, we find that the 5.3--34~$\micron$ optical depth spectrum toward \object{$\zeta$ Oph} is adequately ($\chi_{\nu}^{2}$ = 1.21) simulated by a linear combination of amorphous silicates with Mg$_{2}$SiO$_{4}$, MgFeSiO$_{4}$, and MgSiO$_{3}$ compositions.  Specifically, we find identical results for spectral decomposition models utilizing a third- or fourth-order polynomial flux continuum.  In these cases, the Fe-free and Fe-bearing amorphous silicate species are detected with a significance ranging from 1.6$\sigma$--2$\sigma$.  Furthermore, we find that a fit of equal quality may be obtained for a spectral decomposition model consisting of a second-order polynomial flux continuum.  In this case, the Fe-free and Fe-bearing amorphous silicate components are detected at a similar significance level ranging from 1.7$\sigma$--2.3$\sigma$.  The consistency between the best-fit optical depth results suggests that the amorphous silicate grain composition is not sensitive to the adopted polynomial flux continuum.  Finally, we find no evidence of Fe-bearing amorphous silicates with pyroxene composition (MgFeSi$_{2}$O$_{6}$) or H$_{2}$O ice along the line of sight.

Depending on the choice of the adopted polynomial flux continuum, our analysis indicates that the peak silicate optical depth toward \object{$\zeta$ Oph} is $\tau_{\rm{sil}}$ = 0.031 $\pm$ 0.011 or 0.032 $\pm$ 0.009.  In the local-diffuse ISM, the peak silicate optical depth is correlated with the extinction along the line of sight according to $A_{V}$/$\tau_{\rm{sil}}$ $\approx$ 18.5 \citep{Roche84}.  Adopting a total-to-selective extinction ratio of $R_{V}$ = $A_{V}$/$E_{B-V}$ = 2.55 $\pm$ 0.24 and a color excess of $E_{B-V}$ = 0.32 $\pm$ 0.04 from \citet{Valencic04}, we predict a peak silicate optical depth of $\tau_{\rm{sil}}$ = 0.044 $\pm$ 0.007 for the \object{$\zeta$ Oph} sightline, which is a factor of $\sim$1.4 greater than our measured values.  However, given the uncertainties associated with the measured and predicted values, we conclude that reasonable agreement is found among these quantities.  If the small difference between the measured and predicted values is real, then it might be explained by partial masking of the observed silicate absorption by emission from a warm dust component near \object{$\zeta$ Oph}.

\subsubsection{Degeneracy Test for Silicate Components}

The robustness of the best-fit results was investigated by randomizing the initial values of the parameters 10$^{3}$ times.  On each occasion, we find that the spectral decomposition models converge to the same results.  To further explore the possibility of degeneracy among the best-fit amorphous silicate components, the correlation coefficient, $r$, for each off-diagonal element is computed using
\begin{equation}\label{eq:degen}
r_{ij} = \frac{\mathrm{cov}_{ij}}{\sigma_{i}\sigma_{j}},
\end{equation}
\noindent where $\sigma_{i}$ and $\sigma_{j}$ are the formal parameter uncertainties of the $i$th and $j$th components, respectively, and cov$_{ij}$ is the covariance between the component pairs.  Highly degenerate component pairs are anti-correlated and possess a correlation coefficient near $r = -1$.  Similarly, non-degenerate component pairs are uncorrelated and exhibit a correlation coefficient near $r = 0$.  The most significant degeneracy is found between the Mg$_{2}$SiO$_{4}$ and MgFeSiO$_{4}$ components of the third- and fourth-order polynomial spectral decomposition models.  For this component pair, we calculate a correlation coefficient of $r = -0.84$.  Conversely, we find that only small degeneracies, corresponding to $r = -0.16$ and $r = -0.32$, exist between the MgSiO$_{3}$:Mg$_{2}$SiO$_{4}$ and MgSiO$_{3}$:MgFeSiO$_{4}$ component pairs, respectively.

\begin{figure}[htp]
\centering
\includegraphics*[width=0.482\textwidth]{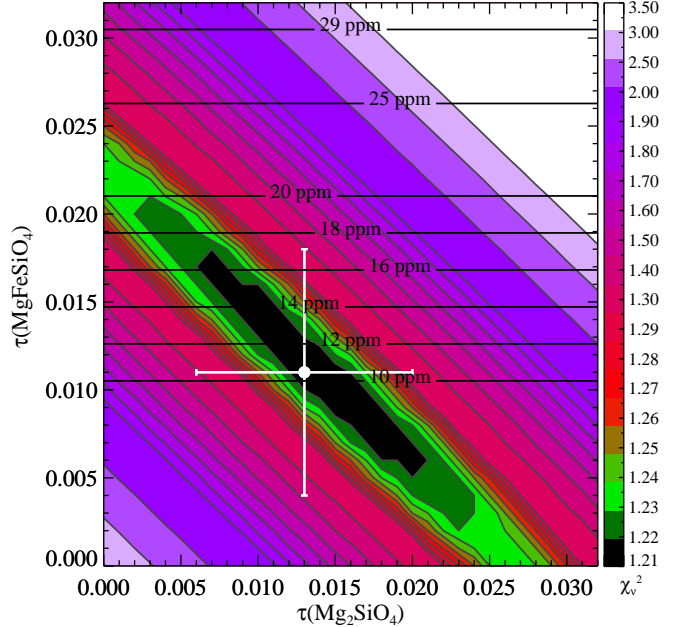}
\caption{Reduced chi-squared ($\chi_{\nu}^{2}$) space of the fourth-order polynomial spectral decomposition model results.  Contours of $\chi_{\nu}^{2}$ for the $\tau$(MgFeSiO$_{4}$) and $\tau$(Mg$_{2}$SiO$_{4}$) parameters are illustrated for values ranging from 1.21 to 3.50.  The shape of the $\chi_{\nu}^{2}$ minimum is characterized by an ellipse with a slope of approximately $-$0.8, indicating that the linear combination of the MgFeSiO$_{4}$ and Mg$_{2}$SiO$_{4}$ components is well determined.  Additionally, the abundance of elemental Fe ($N$(Fe)/$N$(H); black horizontal lines) contained in amorphous silicate grains is also illustrated for each model (see Section \ref{sec:abun} for details of the calculation).  Finally, the best-fit ($\chi_{\nu}^{2}$ = 1.21) optical depth values and formal uncertainties from Table \ref{tab:parameters} (white circle) are indicated for reference.} 
\label{fig:degen}
\end{figure}

In order to visualize the degeneracy between the Fe-free and Fe-bearing amorphous silicate components with olivine stoichometry, we recompute the fourth-order polynomial spectral decomposition model for $33 \times 33$ combinations of the Mg$_{2}$SiO$_{4}$ and MgFeSiO$_{4}$ components.  All model parameters are held at their best-fit values except for the Mg$_{2}$SiO$_{4}$ and MgFeSiO$_{4}$ parameters, which are varied over an optical depth range of 0 $\leq$ $\tau$ $\leq$ 0.032.  Contours of the two parameter $\chi_{\nu}^{2}$ space are presented in Figure \ref{fig:degen} and indicate the following observables: (1) the best-fit ($\chi_{\nu}^{2}$ = 1.21) component pair from Table \ref{tab:parameters} is situated within the innermost contour, which is representative of the true {\it global} minimum.  (2) The $\chi_{\nu}^{2}$ minimum is characterized by an ellipse with a slope of approximately $-$0.8.  (3) The orientation of the $\chi_{\nu}^{2}$ minimum demonstrates that a linear combination of both components is required to accurately simulate the absorption profile by amorphous silicates toward \object{$\zeta$ Oph}.  (4) Although the optical depth of the Mg$_{2}$SiO$_{4}$ and MgFeSiO$_{4}$ components cannot be precisely disentangled from the total line of sight contribution, an indication of their dependence on one another is well determined.  As such, increasing the optical depth of one component and decreasing the other yields a goodness-of-fit of similar quality ($\chi_{\nu}^{2}$ = 1.22).  Hence, the two parameter $\chi_{\nu}^{2}$ space is instrumental in revealing both qualitative and quantitative information that could not otherwise be extracted from the formal 1$\sigma$ parameter uncertainties.

\subsection{H$_{2}$O Ice Mantles}

\subsubsection{Sub-Micron-Sized Grains}

\begin{figure}[htp]
\centering
\includegraphics*[width=0.48\textwidth]{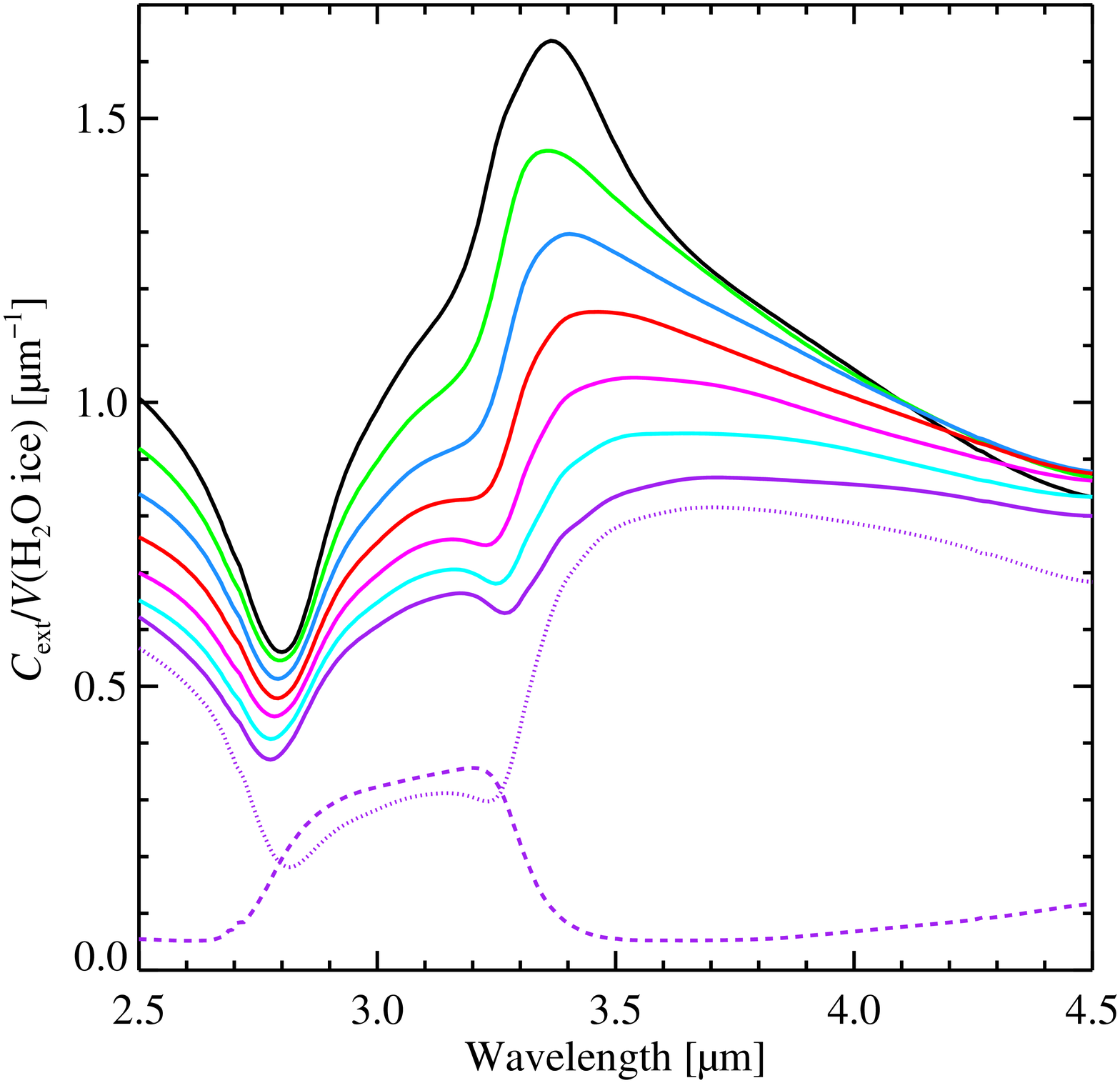}
\caption{Extinction cross sections for grain size distributions of thick H$_{2}$O ice mantles on large grains.  The weighted sum profiles are presented for central grain radii of $a =$ 2.0 (black line), 2.2 (green line), 2.4 (blue line), 2.6 (red line), 2.8 (magenta line), 3.0 (cyan line), and 3.2~$\micron$ (violet line).  The individual scattering (dotted violet line) and absorption (dashed violet line) cross sections are indicated for a grain size distribution with a central radius of $a =$ 3.2~$\micron$.}
\label{fig:H2Oextinction}
\end{figure}

Our model simulations of the \emph{Spitzer}-IRS spectrum toward \object{$\zeta$ Oph} do not require a contribution from the 12.6~$\micron$ H$_{2}$O ice libration feature.  However, because this feature is blended with the strong vibrational modes of amorphous silicates, it may be difficult to detect along lines of sight where H$_{2}$O ice is not abundant.  In contrast, the 3.1~$\micron$ O$\sbond$H stretching mode is intrinsically stronger than the libration mode ($\tau_{3.1}/\tau_{12.6} \approx$ 4.5) and far less affected by overlap with other absorption features, and thus is a more reliable and sensitive tracer of H$_{2}$O ice mantles on sub-micron-sized grains.

Following a similar methodology as before, a spectral decomposition analysis of the \emph{ISO}-SWS spectrum was performed over the 2.8--3.8~$\micron$ wavelength range using only the calculated absorption cross sections of H$_{2}$O ice from Section \ref{sec:specdecomp}.  However, we find no evidence of the 3.1~$\micron$ H$_{2}$O ice feature at the $\geq$1$\sigma$ significance level.  Consequently, from the standard error estimate between the observed optical depth spectrum and the second-order polynomial spectral decomposition model over the 2.8--3.3~$\micron$ spectral region, we deduce a limiting optical depth of $\tau \leq 0.008$ for the 3.1~$\micron$ H$_{2}$O ice feature.  We note that this result corresponds to a limiting optical depth of $\tau \leq 0.002$ for the 12.6~$\micron$ H$_{2}$O ice libration mode, which is a factor of $\sim$2--3 less than the uncertainties reported for best-fit optical depth values of the amorphous silicate components.  Furthermore, the non-detection of H$_{2}$O ice toward \object{$\zeta$ Oph} is in accord with the finding that surfaces of small (sub-micron-sized) grains begin to accumulate H$_{2}$O ice mantles at column densities corresponding to silicate optical depths of $\tau_{\rm{sil}} \gtrsim 0.15 \pm 0.03$ toward several prototypical dark clouds \citep[][and references therein]{Whittet13}.

\subsubsection{Very Large Grains}

\begin{figure}[htp]
\centering
\includegraphics*[width=0.48\textwidth]{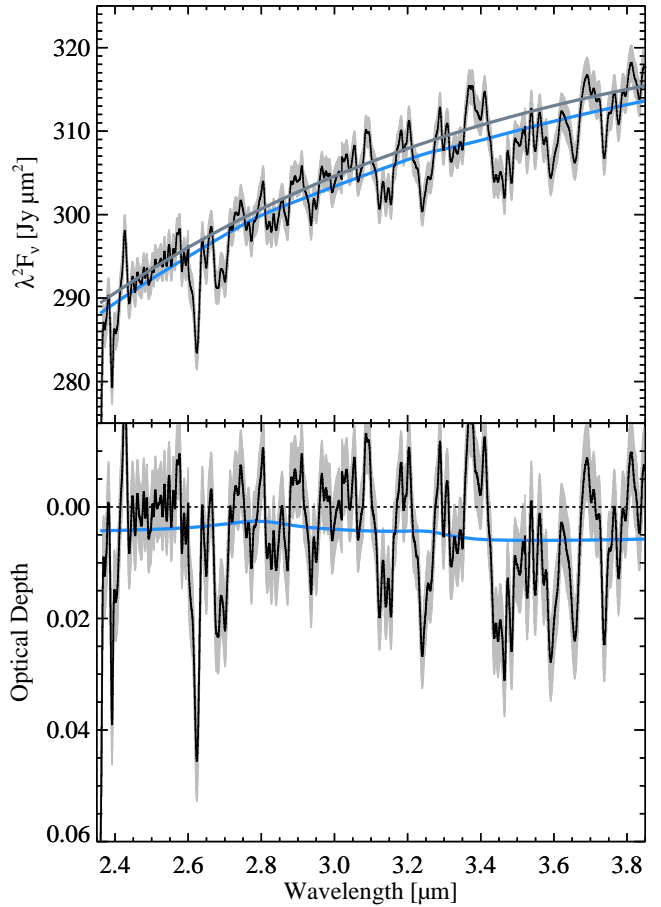}
\caption{\emph{ISO}-SWS 2.4--3.9~$\micron$ spectrum and best-fit second-order polynomial spectral decomposition model for the $\zeta$ Oph sightline.  Top panel: observed spectrum (black line), model flux continuum (gray line), and best-fit model spectrum (blue line).  Bottom panel: observed optical depth spectrum (black line) and best-fit model optical depth (blue line).  The total extinction is modeled by a grain size distribution of thick H$_{2}$O ice mantles with a central radius of $a =$ 2.8~$\micron$.}
\label{fig:H2Omodel}
\end{figure}

Having found no convincing evidence of sub-micron-sized icy grains, we next consider the possibility of thick H$_{2}$O ice mantles on very large grains ($a$ $\gtrsim$ 1~$\micron$).  The existence of micron-sized interstellar grains was proposed by \citet{Greenberg74}.  More recently, possible evidence for the presence of micron-sized interstellar grains have been found by \citet{Wang14}, who modeled the ``flat'' mid-infrared extinction toward a variety of interstellar environments using a population of micron-sized grains.  Furthermore, the recent analysis of three possibly interstellar dust grains returned by the Stardust spacecraft provides tentative evidence for the presence of micron-sized grains in the ISM \citep{Westphal14}.
 
\setlength{\tabcolsep}{0.042 in}
\begin{deluxetable*}{ccccccccc}
\centering
\tablewidth{0pt}
\tablecolumns{9}

\tablecaption{Solid-State Column Densities Toward $\zeta$ Oph}

\tablehead{
\colhead{Poly.\tablenotemark{a}} &
\colhead{$\Sigma$(sil)} &
\colhead{$\Sigma$(Mg$_{2}$SiO$_{4}$)} &
\colhead{$\Sigma$(MgFeSiO$_{4}$)} &
\colhead{$\Sigma$(MgSiO$_{3}$)} &
\colhead{$\Sigma$(Fe$_{3}$O$_{4}$)\tablenotemark{b}} &
\colhead{$N$(H$_{2}$O)} &
\colhead{$N$(POM)\tablenotemark{c}} & 
\colhead{$X$(oli)\tablenotemark{d}} \vspace{0.02 in} \\
\colhead{Degree} &
\colhead{(10$^{-5}$ g cm$^{-2}$)} &
\colhead{(10$^{-6}$ g cm$^{-2}$)} &
\colhead{(10$^{-6}$ g cm$^{-2}$)} &
\colhead{(10$^{-6}$ g cm$^{-2}$)} &
\colhead{(10$^{-6}$ g cm$^{-2}$)} &
\colhead{(10$^{16}$ cm$^{-2}$)} &
\colhead{(10$^{15}$ cm$^{-2}$)} &
\colhead{(\%)} \vspace{0.025 in}\\
\cline{1-9}\vspace{-0.06 in} \\
\multicolumn{9}{c}{\emph{Spitzer}-IRS spectral decomposition models}}
\startdata
2 \dotfill & 1.1 $\pm$ 0.3 & 5.6 $\pm$ 2.4 & 3.8 $\pm$ 2.3 & 1.9 $\pm$ 0.8  & $\leq$2.5 &\nodata & $\leq$9.7 &  85 \\
3 \dotfill & 1.1 $\pm$ 0.4 & 5.2 $\pm$ 2.8 & 4.2 $\pm$ 2.7 & 1.6 $\pm$ 0.8  & $\leq$2.5 & \nodata & $\leq$9.7 &  85 \\
4 \dotfill & 1.1 $\pm$ 0.4 & 5.2 $\pm$ 2.8 & 4.2 $\pm$ 2.7 & 1.6 $\pm$ 0.8 & $\leq$2.5 & \nodata & $\leq$9.7 & 85\vspace{-0.03 in}\\
\cutinhead{\emph{ISO}-SWS spectral decomposition models}\vspace{-0.09 in}\\
2 \dotfill & \nodata & \nodata & \nodata & \nodata  & \nodata & $\leq$1.2\tablenotemark{e} & \nodata & \nodata \\
2 \dotfill & \nodata & \nodata & \nodata & \nodata  & \nodata & $\leq$3.2\tablenotemark{f} & \nodata & \nodata\vspace{0.015 in}
\enddata
\tablecomments{Uncertainties are based on statistical errors in the \emph{Spitzer}-IRS and \emph{ISO}-SWS spectra only.}
\tablenotetext{a}{Degree of the polynomial flux continuum.}
\tablenotetext{b}{Based on the 16~$\micron$ Fe$_{3}$O$_{4}$ optical depth.}
\tablenotetext{c}{Based on the 8.9~$\micron$ POM optical depth.}
\tablenotetext{d}{Mass fraction of amorphous silicates with olivine composition.}
\tablenotetext{e}{Based on the 3.1~$\micron$ H$_{2}$O ice optical depth of sub-micron-sized mantled grains.}
\tablenotetext{f}{Based on the 95\% upper confidence limit of the 3.53~$\micron$ H$_{2}$O ice optical depth for thick mantles with central radius of $a =$ 2.8~$\micron$.}

\label{tab:columns}
\end{deluxetable*}

Adopting amorphous silicate and H$_{2}$O ice ($T = 25$~K) optical constants from \citet{Draine84} and \citet{Mastrapa09}, respectively, the optical properties of large, ice-coated spherical grains are modeled using an extension of Mie theory \citep{Bohren83} for particle radii in the range of 1~$\micron \leq a \leq 4~\micron$.  We assume that one-eighth of the total particle volume is occupied by the amorphous silicate core.  In this thick mantle limit, we note that the ice-coated extinction profile is similar to that of pure homogeneous H$_{2}$O ice spheres.  Finally, to smooth oscillatory structure in the calculated absorption cross sections, we sum over three grain radii of $a-0.2~\micron$, $a$, and $a+0.2~\micron$, with relative weights of 25\%, 50\%, and 25\%, respectively.

The extinction cross sections for grain size distributions of thick H$_{2}$O ice mantles are presented in Figure \ref{fig:H2Oextinction} with central radii of $a =$ 2.0, 2.2, 2.4, 2.6, 2.8, 3.0, and 3.2~$\micron$.  The contributions from absorption and scattering are separately shown for the largest grain size distribution, and demonstrate that the extinction cross section of large grains are dominated by scattering at wavelengths shortward of 2.8 $\micron$ and longward of 3.3~$\micron$.  Moreover, we find that the extinction profile shifts longward and decreases in amplitude with increasing grain radius.

Employing a similar approach as before, a spectral decomposition analysis of the \emph{ISO}-SWS spectrum was performed over the 2.5--3.9~$\micron$ wavelength range using only the aforementioned extinction spectra of very large, spherical H$_{2}$O ice mantles.  In general, we find that the \emph{ISO}-SWS spectrum can be adequately ($\chi_{\nu}^{2} =$ 2.34) simulated using a second-order polynomial flux continuum and grain size distributions with central radii in the range of $2~\micron \leq a \leq 3.2~\micron$.  The most significant (3$\sigma$) evidence of large H$_{2}$O ice mantles is found for central radii of $a =$ 2.6 and 2.8~$\micron$, which result in H$_{2}$O ice optical depths of $\tau =$ 0.006 $\pm$ 0.002 at 3.46 and 3.53~$\micron$, respectively.  The best-fit spectral decomposition model result is shown in Figure \ref{fig:H2Omodel} for a grain size distribution with a central radius of $a =$ 2.8~$\micron$.  We note that the 95\% upper confidence limit of $\tau \leq 0.009$ would only weakly affect the extinction law near 3.5~$\micron$ (i.e., $\Delta A_{\lambda}/E_{B-V} = 1.086 \times (0.009/0.32) = 0.03$).  This result demonstrates that the presence of H$_{2}$O ice in the diffuse ISM may be revealed through scattering by thick icy mantles on large grains.  The notion that such material might form within dense molecular clouds and survive subsequent dispersal to more diffuse regions of the ISM may be supported by the longer lifetimes predicted for thick H$_{2}$O ice mantles \citep[e.g.,][]{Oberg09}.  

\section{Solid-State Elemental Abundances}\label{sec:abun}

The results from the spectral decomposition analysis allow constraints to be placed on the abundances of elements contained within the relevant solids toward \object{$\zeta$ Oph} (i.e., those with vibrational modes in the spectral range covered by the available observations).  A comparison between gas-phase and solid-state abundances in this line of sight may be informative, particularly in regard to the question of any potential ``missing" reservoir of interstellar O, as discussed in Section \ref{sec:intro}.

In order to estimate the solid-state elemental abundances of Mg, Si, O, and Fe relative to H, we adopt a total hydrogen column density of $N(\rm{H}) = 1.42 \times 10^{21}$ cm$^{-2}$ for the diffuse clouds toward \object{$\zeta$ Oph} \citep{Savage77}.  However, the gas-phase interstellar absorption along this line of sight is known to originate from two separate diffuse cloud components \citep[e.g.,][]{Hobbs73, Snow79, Savage92}.  The high-velocity component ($v_{\odot} = -27$ km s$^{-1}$) samples only the modest depletion associated with lower density clouds.  In contrast, the low-velocity component ($v_{\odot} = -15$ km s$^{-1}$) samples the highly depleted, denser cloud medium, and accounts for most of the observed hydrogen along the line of sight \citep{Jenkins09}.  Adopting a synthetic hydrogen column density of $N(\rm{H}) = 3.39 \times 10^{19}$ cm$^{-2}$ for the high-velocity component from \citet{Jenkins09}, we estimate a hydrogen column density of $N(\rm{H}) = 1.39 \times 10^{21}$ cm$^{-2}$ for the low-velocity component, which accounts for 97.6\% of the total hydrogen column density toward \object{$\zeta$~Oph}.

\setlength{\tabcolsep}{0.103 in}
\begin{deluxetable*}{lcccccccccccc}
\centering
\tablecolumns{13}

\tablecaption{Solid-State Elemental Abundances Toward $\zeta$ Oph}

\tablehead{
\colhead{} & 
\multicolumn{4}{c}{Amorphous Silicates} &
\colhead{} &
\multicolumn{2}{c}{Fe$_{3}$O$_{4}$} &
\colhead{} &
\multicolumn{2}{c}{H$_{2}$O Ice Mantles\tablenotemark{b}} &
\colhead{} &
\colhead{POM} \vspace{0.03 in} \\ 
\cline{2-5} \cline{7-8} \cline{10-11} \cline{13-13} \vspace{-0.05 in} \\ 
\colhead{Poly.\tablenotemark{a}} &
\colhead{$\frac{N(\rm{Mg})}{N(\rm{H})}$} &
\colhead{$\frac{N(\rm{Si})}{N(\rm{H})}$} &
\colhead{$\frac{N(\rm{O})}{N(\rm{H})}$} &
\colhead{$\frac{N(\rm{Fe})}{N(\rm{H})}$} &
\colhead{} &
\colhead{$\frac{N(\rm{O})}{N(\rm{H})}$} &
\colhead{$\frac{N(\rm{Fe})}{N(\rm{H})}$} &
\colhead{} &
\colhead{$\frac{N(\rm{O})}{N(\rm{H})}$} &
\colhead{$\frac{N(\rm{O})}{N(\rm{H})}$} &
\colhead{} &
\colhead{$\frac{N(\rm{O})}{N(\rm{H})}$} \vspace{0.025 in}\\
\colhead{Degree} &
\colhead{(ppm)} &
\colhead{(ppm)} &
\colhead{(ppm)} &
\colhead{(ppm)} &
\colhead{} &
\colhead{(ppm)} &
\colhead{(ppm)} &
\colhead{} &
\colhead{(ppm)} &
\colhead{(ppm)} &
\colhead{} &
\colhead{(ppm)} \vspace{0.025 in}\\
\cline{1-13}\vspace{-0.06 in} \\
\multicolumn{13}{c}{\emph{Spitzer}-IRS spectral decomposition models} }
\startdata
2 \dotfill & 52 $\pm$ 16 & 34 $\pm$ 10 & 130 $\pm$ 39 & 10 $\pm$ 6 & & $\leq$19 & $\leq$14 & & \nodata & \nodata & & $\leq$7 \\
3 \dotfill & 49 $\pm$ 19 & 33 $\pm$ 11 & 126 $\pm$ 45 & 10 $\pm$ 7 & & $\leq$19 & $\leq$14 & &\nodata & \nodata & & $\leq$7 \\
4 \dotfill & 49 $\pm$ 19 & 33 $\pm$ 11 & 126 $\pm$ 45 & 10 $\pm$ 7 & & $\leq$19 & $\leq$14 & & \nodata & \nodata & & $\leq$7\vspace{-0.03 in} \\
\cutinhead{\emph{ISO}-SWS spectral decomposition models}\vspace{-0.09 in}\\
2 \dotfill & \nodata &  \nodata  &  \nodata &  \nodata  &  & \nodata & \nodata & & $\leq$9 & $\leq$23 & & \nodata\vspace{-0.03 in}\\
\cutinhead{Elemental abundances absent from the gas-phase\tablenotemark{c}}
Protosolar\tablenotemark{d} & 42 $\pm$ 4\phantom{1} & 34 $\pm$ 3\phantom{1} & 230 $\pm$ $^{72}_{65}$ & 35 $\pm$ 3\phantom{1} & & \nodata & \nodata & & \nodata & \nodata & & \nodata\vspace{0.025 in}\\
Protosolar$+$GCE\tablenotemark{e} & 46 $\pm$ $^{5\phantom{1}}_{4\phantom{1}}$ & 41 $\pm$ 4\phantom{1} & 282 $\pm$ $^{78}_{71}$ & 48 $\pm$ $^{5\phantom{1}}_{4\phantom{1}}$ & & \nodata & \nodata & & \nodata & \nodata & & \nodata\vspace{0.025 in}\\
F and G stars\tablenotemark{f} & 42 $\pm$ $^{27}_{17}$ & 39 $\pm$ $^{29}_{17}$ & \phantom{1}282 $\pm$ $^{207}_{155}$ & 41 $\pm$ $^{32}_{18}$ & & \nodata & \nodata & & \nodata & \nodata & & \nodata\vspace{0.025 in}\\
B stars\tablenotemark{g} & 34 $\pm$ 4\phantom{1} & 30 $\pm$ 4\phantom{1} & 268 $\pm$ 72 & 33 $\pm$ 2\phantom{1} & & \nodata & \nodata & & \nodata & \nodata & & \nodata\vspace{0.015 in}
\enddata
\vspace{-0.03cm}
\tablecomments{Uncertainties are based on statistical errors in the \emph{Spitzer}-IRS and \emph{ISO}-SWS spectra only.  Abundances are calculated assuming a hydrogen column density of $N$(H) = 1.39 $\times 10^{21}$ cm$^{-2}$ for the low-velocity ($v_{\odot} = -15$ km s$^{-1}$) component.}
\tablenotetext{a}{Degree of the polynomial flux continuum.}
\tablenotetext{b}{Abundance of elemental O in sub-micron-sized and very large H$_{2}$O ice mantles, respectively.  The latter is derived from the 95\% upper confidence limit of the 3.53~$\micron$ H$_{2}$O ice optical depth for thick mantles with central radius of $a =$ 2.8~$\micron$.}
\tablenotetext{c}{Based on the difference between reference abundances and the observed gas-phase abundances derived from \citet{Jenkins09}: 1.9 $\pm$ 0.1~ppm (Mg), 1.6 $\pm$ 0.1~ppm (Si), 307 $\pm$ 30~ppm (O), and 0.18 $\pm$ 0.01~ppm (Fe).}
\tablenotetext{d}{\citet{Asplund09}}
\tablenotetext{e}{\citet{Asplund09, Chiappini03}}
\tablenotetext{f}{\citet{Bensby05, Lodders09}}
\tablenotetext{f}{\citet{Nieva12}}
\label{tab:abundances}
\end{deluxetable*}

\subsection{Amorphous Silicates}

Under the assumption that the absorption features in the \emph{Spitzer}-IRS spectrum are entirely attributed to dust located in the highly depleted, low-velocity component, the relative abundance of element $x$ contained in silicates is determined following the prescription of \citet{Whittet90}:
\begin{equation}\label{eq:sil}
\frac{N(x)}{N(\rm{H})} = \frac{1}{N(\rm{H})} \sum_{i} \frac{f_{x,i} \tau_{\lambda,i}}{\mu_{x}\langle\kappa_{\lambda,i}\rangle}, 
\end{equation}
\noindent where $N(\rm{H}) = 1.39 \times 10^{21}$ cm$^{-2}$, $f_{x,i}$ is the mass fraction of element $x$ in the $i$th silicate component, $\mu_{x}$ is the atomic mass of element $x$, $\tau_{\lambda,i}$ is the observed peak optical depth of the $i$th silicate component, and 
\begin{equation}
\langle\kappa_{\lambda,i}\rangle = \frac{\langle C_{\mathrm{abs},i}\rangle}{V_{i}\rho_{i}}
\end{equation}
\noindent is the shape-averaged peak mass absorption coefficient for the $i$th silicate component with a shape-averaged absorption cross section per unit volume $\langle$$C_{\mathrm{abs},i}$$\rangle/V_{i}$ and specific density $\rho_{i}$.  Using the laboratory-measured specific densities of $\rho =$ 3.2, 3.71, and 2.71 g cm$^{-3}$ for Mg$_{2}$SiO$_{4}$, MgFeSiO$_{4}$, and MgSiO$_{3}$ \citep{Henning96, Dorschner95}, we calculate shape-averaged peak mass absorption coefficients of $\langle\kappa_{\lambda}\rangle =$ 2510, 2624, and 4853 cm$^{2}$ g$^{-1}$, respectively.

The total and individual mass column densities, $\Sigma = \tau_{\lambda}/\langle\kappa_{\lambda}\rangle$, of amorphous silicates, as well as the mass fraction of amorphous silicates with olivine composition, are presented in Table~\ref{tab:columns}.  The solid-state elemental abundances of Mg, Si, O, and Fe are listed in Table~\ref{tab:abundances}.  For comparison, reference elemental dust abundances, based on average abundances in young F and G stars \citep{Bensby05, Lodders09}, protosolar abundances \citep{Asplund09}, protosolar abundances with Galactic chemical enrichment (GCE) over the past 4.56 Gyr \citep{Asplund09, Chiappini03}, and average abundances in B stars \citep{Nieva12}, are calculated using previously published and revised gas-phase column densities of the low-velocity cloud component toward \object{$\zeta$ Oph} \citep[][and references therein]{Jenkins09}.

In general, we find that solid-state elemental abundances derived from the spectral decomposition models are in good agreement with those predicted from stellar and solar abundances.  Specifically, we find that the elemental abundance of Mg is slightly-to-moderately over-consumed, requiring $\sim$7\%-53\% more than the available abundance.  Similarly, our model results indicate that nearly all ($\sim$80\%--100\%) of the available elemental Si abundance is present in amorphous silicates.  (When considering the average reference abundance in B stars, we note that the elemental abundance of Si is slightly over-consumed by $\sim$10\%--13\%.)  In contrast to the substantial fraction of elemental Mg and Si contained in amorphous silicates, we find that $\sim$70\%--79\% of the available elemental Fe abundance is expected to be present in some other solid-state species.  Finally, our model results indicate that only $\sim$45\%--57\% of the elemental O that is missing from the gas-phase is present in amorphous silicate grains.  Independent of the adopted reference abundances, these results suggest that additional reservoirs of interstellar Fe and O are needed along the line of sight toward \object{$\zeta$ Oph}.

\begin{figure}[htp]
\centering
\includegraphics*[width=0.48\textwidth]{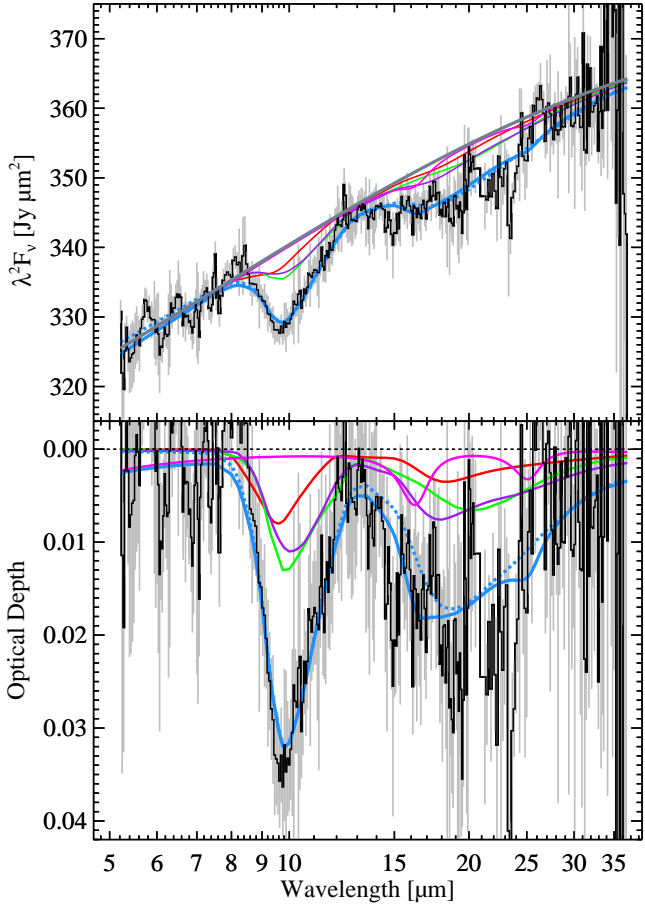}
\caption{\emph{Spitzer}-IRS spectrum and best-fit fourth-order polynomial spectral decomposition model for the $\zeta$ Oph sightline.  Top panel: observed spectrum (black line), model flux continuum (gray line), and best-fit model spectrum with (solid blue line) and without (dotted blue line) Fe$_{3}$O$_{4}$.  Bottom panel: observed optical depth spectrum (black line) and best-fit model optical depth with (solid blue line) and without (dotted blue line) Fe$_{3}$O$_{4}$.  The total absorption is modeled as a linear combination of Fe$_{3}$O$_{4}$ (magenta line) and amorphous silicates with Mg$_{2}$SiO$_{4}$ (green line), MgFeSiO$_{4}$ (violet line), and MgSiO$_{3}$ (red line) stoichiometries.  A limiting optical depth of $\tau$ $\leq$ 0.006 for the 16~$\micron$ Fe$_{3}$O$_{4}$ feature is assumed.}
\label{fig:Fe3O4Limit}
\end{figure}

\subsection{Additional Repositories of Interstellar Fe}

\subsubsection{Fe-rich Amorphous Silicates}

Similar to previous diffuse ISM studies \citep[e.g.,][]{Kemper04, Chiar06, Min07}, Fe-bearing amorphous silicates with equal amounts of elemental Mg and Fe have been utilized to simulate the absorption profile of amorphous silicates toward \object{$\zeta$ Oph}.  To investigate whether a significant fraction of interstellar Fe may be present in amorphous silicates with Fe-to-Mg ratios greater than unity (Fe/Mg $>$ 1), the \emph{Spitzer}-IRS spectrum toward \object{$\zeta$ Oph} is further modeled using the Mg$_{4}$Fe$_{6}$Si$_{5}$O$_{20}$ species instead of the MgFeSiO$_{4}$ component.  However, we find that the slightly Fe-rich component is not utilized by the nonlinear least-squares fitting routine; this is true for all (2nd-, 3rd-, and 4th-order) polynomial flux continua considered in Section \ref{sec:specdecomp}.  This result is likely a consequence of the broadened spectral profiles possessed by Fe-rich amorphous silicates \citep{Dorschner95}, and further demonstrates that stoichiometries with Fe/Mg $\lesssim$ 1 are required when simulating the absorption profile of diffuse ISM silicates.  Finally, if we assume the unlikely notion that MgFeSiO$_{4}$ is the only olivine-like species present toward \object{$\zeta$ Oph}, we find that the abundance of elemental Fe is increased by only $\sim$10~ppm, as indicated in Figure \ref{fig:degen}.  Therefore, we conclude that some other Fe-bearing compound must account for the $\sim$23--38~ppm of elemental Fe missing along the line of sight.

\subsubsection{Fe Oxides and Metallic Fe}

In addition to its presence in silicate grains, interstellar Fe could exist in pure metallic Fe or Fe-bearing oxide particles of FeO, $\gamma$-Fe$_{2}$O$_{3}$, and Fe$_{3}$O$_{4}$ \citep[e.g.,][and references therein]{Draine13}.  Metallic Fe possesses no active vibrational modes in the mid-infrared wavelength region and contributes only to the continuum opacity of dust.  Pure metallic Fe particles have been found toward the circumstellar dust shells surrounding asymptotic giant branch stars \citep[e.g.,][]{Kemper02}, but may be unlikely to remain pristine in the oxygen-rich environment of the diffuse ISM \citep{Jones90}.  In contrast, Fe-bearing oxides are considered potential constituents of interstellar grains \citep{Sofia94}.  Depending on the precise composition of these species, Fe-bearing oxides exhibit single or multiple vibrational resonances within the 17--28~$\micron$ spectral region \citep{Henning95, Glotch06, Glotch09}.  However, because their spectral features are blended with the strong bending mode of amorphous silicates, Fe-bearing oxides have not been detected in the diffuse ISM, and thus are not usually considered to be a significant component of interstellar dust \citep[e.g.,][]{Chiar06}.  More recently, \citet{Li13} found that nanoparticles of FeO are not likely the carrier species responsible for the 21~$\micron$ emission observed toward post-asymptotic giant branch stars. 

To examine whether Fe-bearing oxides are a major reservoir of elemental Fe toward \object{$\zeta$ Oph}, we consider irregularly shaped particles of FeO (w\"{u}stite) and Fe$_{3}$O$_{4}$ (magnetite) as potential line of sight absorbers.  While $\gamma$-Fe$_{2}$O$_{3}$ (maghemite) is known to possesses broad resonance features near 17 and 27 $\micron$, its absorption opacity contributes primarily to the featureless dust continuum, and a non-detection of the features does not provide a useful upper limit estimate on the abundance of $\gamma$-Fe$_{2}$O$_{3}$.  For this reason, we do not include $\gamma$-Fe$_{2}$O$_{3}$ in our spectral analysis.

As a result of the low signal-to-noise ratio within the 19--35~$\micron$ region of the \emph{Spitzer}-IRS optical depth spectrum, we are unable to accurately quantify the presence of Fe-bearing oxides along the sightline using the spectral decomposition analysis from Section \ref{sec:specdecomp}.  Alternatively, from the standard error estimate between the observed and modeled optical depth spectrum within the 15--30~$\micron$ wavelength region, we deduce a limiting optical depth of $\tau$ $\leq$ 0.009 for the 22~$\micron$ FeO feature.  Assuming a specific density of $\rho =$ 5.70~g cm$^{-3}$ \citep{Henning95} and a DHS ($f_{\rm{max}}$ = 0.7) shape-averaged peak absorption cross section of $\langle C_{\mathrm{abs}}\rangle/V =$ 13590~cm$^{-1}$, we set a limit of $\leq$23~ppm for the abundance of elemental Fe in FeO grains.  (In this context, a limiting abundance of $\leq$23~ppm for elemental O in FeO is also deduced.)  Nevertheless, we emphasize that the generous limit on the optical depth of FeO, which is likely a consequence of the appreciable noise present in the 19--30~$\micron$ spectral region, would considerably affect the shape of the optical depth spectrum toward \object{$\zeta$ Oph}, and therefore should be considered with caution. 

Similarly, from the standard error estimate between the observed and modeled optical depth spectrum within the 13--18~$\micron$ wavelength region, we calculate a limiting optical depth of $\tau$ $\leq$ 0.006 for the 16~$\micron$ Fe$_{3}$O$_{4}$ feature.  In contrast to the FeO investigation, we note that the limit on the optical depth of Fe$_{3}$O$_{4}$ does not strongly affect the appearance of the 13--30~$\micron$ region of the best-fit models from Section \ref{sec:silcomp}, as illustrated in Figure \ref{fig:Fe3O4Limit}.  Adopting a specific density of $\rho =$ 5.18~g cm$^{-3}$ \citep{Draine13} and a DHS shape-averaged peak absorption cross section of $\langle C_{\mathrm{abs}}\rangle/V =$ 12260~cm$^{-1}$, we deduce limiting abundances of $\leq$14~ppm and $\leq$19~ppm for elemental Fe and O contained in Fe$_{3}$O$_{4}$, respectively.  In summary, these results establish that Fe$_{3}$O$_{4}$ may account for $\lesssim$37--61\% of the missing elemental Fe abundance toward \object{$\zeta$ Oph}.

\subsection{Missing Reservoirs of Interstellar O}

As previously noted, the rate at which the observed depletion of elemental O varies with mean line of sight density in the diffuse ISM is more rapid than can be accounted for by exchange of atoms with amorphous silicates and metallic oxide particles \citep{Jenkins09}.  In consequence, a substantial fraction of interstellar O is thought to reside in an additional reservoir, presumably in combination with some other abundant element(s).  Given that hydrogen and carbon are the only reactive elements with sufficiently large interstellar abundance (and given the improbability of substantial amounts of O$_{2}$ existing in such environments), it follows that H$_{2}$O ice and organic refractory compounds are attractive candidates as carriers of the missing elemental O \citep{Jenkins09, Whittet10}.

\subsubsection{H$_{2}$O Ice}

The abundance of elemental O contained in small H$_{2}$O ice mantles may be directly inferred from its line of sight column density \citep{Allamandola88}:
\begin{equation}
N(\mathrm{H_{2}O}) \approx  \frac{\int \tau_{\tilde{\nu}}d\tilde{\nu}}{A}, 
\end{equation}
\noindent where $A$ is the laboratory-derived intrinsic band strength and $\int \tau_{\tilde{\nu}}d\tilde{\nu}$ is the integrated optical depth in the wavenumber domain, $\tilde{\nu}$.  Adopting an intrinsic band strength of $A =$ 2.0 $\times$ 10$^{-16}$ cm molecule$^{-1}$ \citep{Hagen81} for the 3.1~$\micron$ stretching mode, we calculate a limiting column density of $N$(H$_{2}$O) $\lesssim$ 1.2 $\times$ 10$^{16}$ cm$^{-2}$.  The abundance of elemental O contained in small H$_{2}$O ice mantles toward the low-velocity component of \object{$\zeta$~Oph} is thus no more than 9~ppm (i.e., no more than 6\%--9\% of the missing elemental O abundance).  We note that an even stricter limit ($<$2~ppm) was reported toward the highly reddened ($A_{V} \approx 10$) luminous star \object{Cyg OB2 no.\ 12} \citep{Whittet97}, a sightline also thought to be dominated by diffuse ISM material at comparatively high density \citep[e.g.,][]{Geballe99, Gredel01}.  These results demonstrate that H$_{2}$O ice mantles on sub-micron-sized grains are not a significant reservoir of elemental O near the diffuse-dense ISM transition.

\begin{figure}[htp]
\centering
\includegraphics*[width=0.48\textwidth]{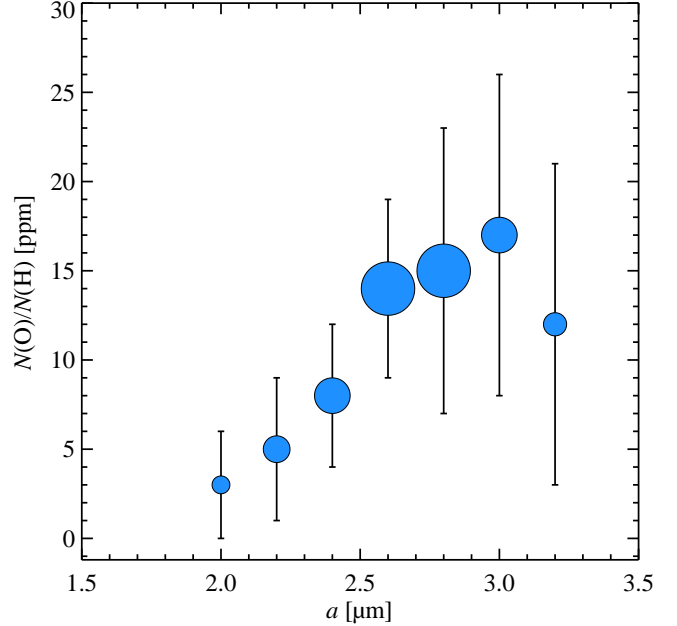}
\caption{Abundance of elemental O that may be contained in thick H$_{2}$O ice mantles on large grains along the $\zeta$ Oph sightline for grain size distributions with central radii of $a =$ 2.0, 2.2, 2.4, 2.6, 2.8, 3.0, and 3.2 $\micron$.  The magnitude of each data point represents the significance of the best-fit H$_{2}$O ice optical depth parameters, which varies over the 1$\sigma$--3$\sigma$ level.}
\label{fig:H2Oabundances}
\end{figure}

By analogy with Equation \ref{eq:sil}, the abundance of elemental O in thick H$_{2}$O ice mantles on large grains is given by 
\begin{equation}
\frac{N(\mathrm{O})}{N(\mathrm{H})} = \frac{1}{N(\rm{H})} \frac{\tau_{\lambda}}{\sigma_{g} Q_{\mathrm{ext},\lambda}} \frac{\rho V_\mathrm{{H_{2}O}}}{18 m_{\mathrm{H}}}, 
\end{equation}
\noindent where $\sigma_{g}$ is the geometric cross section of the grains ($\pi a^{2}$), $V_{\mathrm{H_{2}O}}$ is the volume of the mantle, and $Q_{\rm{ext}, \lambda}$ is the peak extinction efficiency of H$_{2}$O ice.  Assuming a specific density of $\rho =$ 1.1~g~cm$^{-3}$ for amorphous H$_{2}$O ice \citep{Narten76}, the abundance of elemental O in thick H$_{2}$O ice mantles is presented in Figure \ref{fig:H2Oabundances} for grain size distributions with central radii in the range of $2~\micron \leq a \leq 3.2~\micron$.  Considering the 95\% upper confidence limit for a grain size distribution central radius of $a = 2.8~\micron$, we find that as much as $\sim$23~ppm of elemental O may be stored in thick H$_{2}$O ice mantles on large grains toward \object{$\zeta$~Oph}.  Although this result is more than a factor of two larger than the upper limit abundance obtained for small icy grains, it indicates that thick H$_{2}$O mantles on large ($a = 2.8~\micron$) grains can only account for 15\%--23\% of the missing elemental O abundance.   

\subsubsection{Organic Compounds}

We next consider organic refractory matter as a possible repository of interstellar O toward \object{$\zeta$ Oph}.  It is well-known that laboratory experiments simulating the energetic processing (e.g., ultraviolet photolysis or ion bombardment) of interstellar ices lead to the production of organic refractory matter with significant oxygen content \citep[e.g.,][]{Greenberg95, Pendleton02, Munoz03}.  These products include species with carbonyl molecular groups (esters, amides, and carboxylic acid salts) and species related to polyoxymethylene (POM), a polycrystalline material consisting of linear chains of formaldehyde molecules [($\sbond$CH$_{2}$O$\sbond$)$_{n}$].  The ester and amide species possess minor C$\dbond$O stretching modes near 5.7 and 6.0~$\micron$, respectively, while carboxylic acid salts are known to exhibit a strong COO$^{-}$ stretching mode at 6.3~$\micron$ \citep{Munoz03}.  However, apart from the detection of weak $\sim$6~$\micron$ absorption toward carbon-rich Wolf-Rayet stars \citep{Schutte98, Chiar01}, evidence for the presence of these O-bearing organic compounds in the diffuse ISM has remained elusive.

\setlength{\tabcolsep}{0.191 in}
\begin{deluxetable*}{lccccc}
\centering
\tablewidth{0pt}
\tablecolumns{6}

\tablecaption{Summary of the Observed Elemental Budget Toward $\zeta$ Oph}

\tablehead{
\colhead{Material} &
\colhead{$\frac{N(\rm{Mg})}{N(\rm{H})}$} &
\colhead{$\frac{N(\rm{Si})}{N(\rm{H})}$} &
\colhead{$\frac{N(\rm{O})}{N(\rm{H})}$} &
\colhead{$\frac{N(\rm{Fe})}{N(\rm{H})}$} &
\colhead{Reference} \vspace{0.025 in}\\
\colhead{} &
\colhead{(ppm)} &
\colhead{(ppm)} &
\colhead{(ppm)} &
\colhead{(ppm)} &
\colhead{} \vspace{0.01 in}}\\
\startdata
Adopted total \dotfill & 47.9 $\pm$ $^{4.6}_{4.2}$ & 42.7 $\pm$ $^{4.1}_{3.8}$ & 589 $\pm$ $^{72}_{64}$ & 47.9 $\pm$ $^{4.6}_{4.2}$  & 1\vspace{0.025 in}\\
Gas \dotfill & \phantom{1}1.9 $\pm$ 0.1 & \phantom{1}1.6 $\pm$ 0.1 & 307 $\pm$ 30 & \phantom{..}0.18 $\pm$ 0.01 & 2 \\
Silicates \dotfill & \phantom{1}49 $\pm$ 19 & \phantom{1}33 $\pm$ 11 & 126 $\pm$ 45 & 10 $\pm$ 7 & 3\vspace{0.015 in}\\
\cline{1-6}\vspace{-0.07 in}\\
Unidentified \dotfill & \phantom{'}$-$3 $\pm$ $^{20}_{19}$ & \phantom{11}8 $\pm$ 12 & 156 $\pm$ $^{90}_{84}$ & 38 $\pm$ 8 & \vspace{0.030 in}\\
\cline{1-6}\vspace{-0.06 in}\\
H$_{2}$O Ice (sub-micron-sized) \dotfill & \nodata & \nodata & $\leq$9 & \nodata & 3 \\
H$_{2}$O Ice ($a \approx 3$~$\micron$) \dotfill & \nodata & \nodata & \phantom{1}$\leq$23 & \nodata & 3 \\
Fe$_{3}$O$_{4}$ \dotfill & \nodata & \nodata & \phantom{1}$\leq$19 & \phantom{1}$\leq$14 & 3 \\
POM \dotfill & \nodata & \nodata & $\leq$7 & \nodata & 3 \\
Other \dotfill & \phantom{.}$-$3 $\pm$ $^{20}_{19}$ & \phantom{11}8 $\pm$ 12 & (98-156) $\pm$ $^{90}_{84}$\phantom{(98-)} & (24-38) $\pm$ 8\phantom{(24-)} & \vspace{0.015 in}
\enddata
\tablerefs{(1) \citet{Asplund09}, \citet{Chiappini03}; (2) \citet{Jenkins09}; (3) this work.}
\label{tab:summary}
\end{deluxetable*}

\begin{figure}[htp]
\centering
\includegraphics*[width=0.48\textwidth]{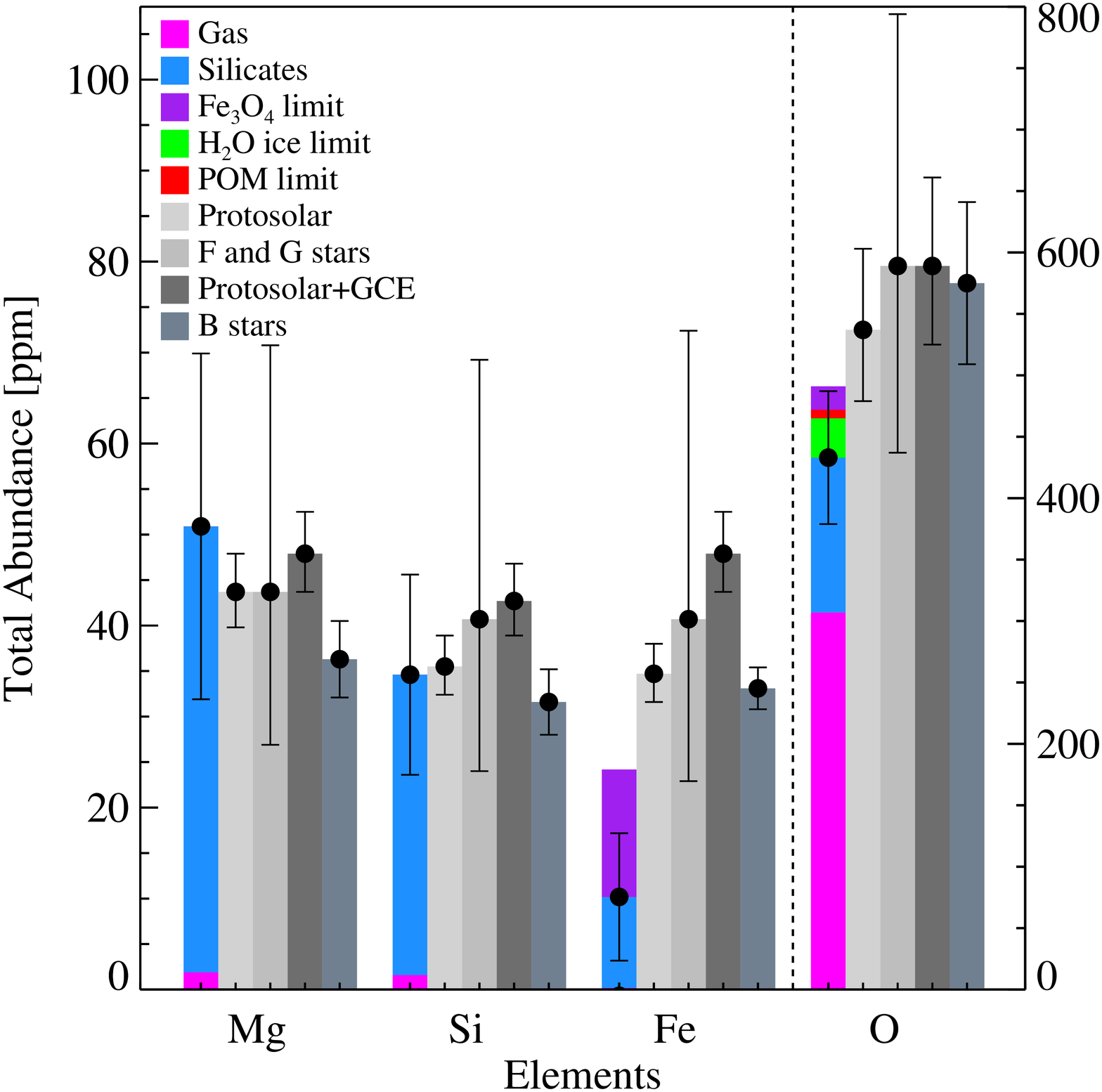}
\caption{Total elemental abundances along the $\zeta$ Oph sightline with respect to reference abundances.  The reference values are based on average abundances in young F and G stars \citep{Bensby05, Lodders09}, protosolar abundances \citep{Asplund09}, protosolar abundances with Galactic chemical enrichment \citep{Asplund09, Chiappini03}, and average abundances in B stars \citep{Nieva12}.  The gas-phase elemental abundances of Mg, Si, O, and Fe (magenta) are based on measured column densities of the low-velocity ($v_{\odot}$ $= -$15 km s$^{-1}$) component \citep[][and references therein]{Jenkins09}.  The solid-state abundances (blue) are representative of the total number of atoms contained in amorphous silicate grains, and are derived from the fourth-order polynomial spectral decomposition model.  Upper limit abundance estimates are indicated for elements contained in H$_{2}$O ice mantles (green), Fe$_{3}$O$_{4}$ (violet), and POM (red).  The former represents the abundance of elemental O that may be contained in H$_{2}$O ice mantles on both sub-micron-sized ($\leq$9~ppm) and very large ($\leq$23~ppm) grains.}
\label{fig:abundances}
\end{figure}

The quality of the 5--8~$\micron$ region of the \emph{Spitzer}-IRS optical depth spectrum toward \object{$\zeta$ Oph} does not permit any meaningful profile comparison with laboratory spectra of carbonyl molecular groups.  In consequence, we make no attempt to constrain the possible fraction of interstellar O contained in carbonyl groups, and further consider the possibility of POM-like species instead.

To explore the presence of POM-like species in the diffuse ISM, we adopt optical constants derived from the laboratory spectrum of pure POM \citep{Whittet76}.  The optical properties of POM are simulated using the DHS grain shape model ($f_{\rm{max}}$ = 0.7) in the Rayleigh limit.  We find that the POM species possesses a strong C$\sbond$O$\sbond$C vibrational mode at 8.95~$\micron$, as well as several weaker features at 6.82, 7.34, 8.05, 11.36, and 11.80~$\micron$.  Following a similar methodology as before, a spectral decomposition analysis of the \emph{Spitzer}-IRS spectrum is performed using the absorption spectra of amorphous silicates and POM.  However, we find no evidence of the 8.9~$\micron$ POM feature at the $\geq$1$\sigma$ significance level.  Alternatively, from the standard error estimate between the observed and modeled optical depth spectrum over the 8.5--9.1~$\micron$ wavelength region, we calculate a limiting optical depth of $\tau$ $\leq$ 0.002 for the 8.9~$\micron$ POM feature.  Adopting an intrinsic band strength of $A =$ 9.7 $\times$ 10$^{-18}$ cm molecule$^{-1}$ \citep{Schutte93}, the limiting column density of POM toward this diffuse sightline is $N$(POM) $\lesssim$ 9.7 $\times$ 10$^{15}$~cm$^{-2}$.  Thus, we conclude that no more than $\leq$7~ppm of elemental O is expected to be found in POM along the \object{$\zeta$ Oph} line of sight.   

\section{Discussion}\label{sec:discuss}

We have shown that the 5--36~$\micron$ \emph{Spitzer}-IRS spectrum toward \object{$\zeta$ Oph} can be adequately simulated using an ensemble of small, irregularly shaped amorphous silicate grains of olivine (Mg$_{2}$SiO$_{4}$ and MgFeSiO$_{4}$) and pyroxene (MgSiO$_{3}$) compositions.  The average stoichiometry of the grains corresponds to Mg/Si $=$ 1.5, O/Si $=$ 3.8, and Fe/Si $=$ 0.3, indicating that the amorphous silicate mass fraction toward the diffuse sightline is dominated by olivine-like compositions ($X_{\rm{oli}}$ $\approx$ 85\%).  We note that this result is in agreement with those of \citet{Min07} and \citet{vanBreemen11}, who found that amorphous silicates in the diffuse ISM are Mg-rich and primarily olivine-like in composition.

Presolar silicate grains found in primitive meteorites possess both stoichiometric (i.e., pyroxene- and olivine-like stoichiometries) and non-stoichiometric compositions, with a wide range of Mg/Fe ratios \citep{Nguyen10, Bose12}.  These grains, identified by virtue of isotopic anomalies, are ``stardust'', condensed in stellar outflows.  Approximately 20\%--25\% of stardust grains found in meteorites are crystalline \citep[][and references therein]{Nguyen10}, and are thought to have undergone only minimal processing during passage through the diffuse ISM \citep[e.g.,][]{Vollmer09}.  In contrast, only $\lesssim$2\% of interstellar silicates are of crystalline nature \citep{Kemper05}, implying that stardust grains comprise at most $\sim$10\% of interstellar silicate material \citep{Draine09}.  Nevertheless, the varied silicate compositions found among stardust grains are not in conflict with our findings on the bulk composition of diffuse ISM silicates toward \object{$\zeta$ Oph}.

Upon further investigation, we find no conclusive evidence for the presence of Fe-bearing oxides, POM species, or H$_{2}$O ice mantles on small grains along the line of sight.  However, from the spectral decomposition analysis of the \emph{ISO}-SWS spectrum, we find tentative evidence for the presence of thick H$_{2}$O ice mantles on large grains ($a = 2.8~\micron$).  These results establish that amorphous silicate grains account for at least 126 $\pm$ 45~ppm of elemental O toward \object{$\zeta$ Oph}, while Fe$_{3}$O$_{4}$, POM species, and H$_{2}$O ice mantles on sub-micron-sized grains can only account for $\leq$19~ppm, $\leq$7~ppm, and $\leq$9~ppm of elemental O, respectively, and therefore are not significant repositories of interstellar O along this line of sight.  Furthermore, tentative evidence for thick H$_{2}$O ice mantles on large grains suggests that $\leq$23~ppm of elemental O may be present in icy grain mantles with radii $a = 2.8~\micron$, thereby establishing that H$_{2}$O ice mantles may collectively contain $\leq$32~ppm of elemental O. 

The total elemental abundance of Mg, Si, O, and Fe in the low-velocity component ($v_{\odot} = -15$~km~s$^{-1}$) toward \object{$\zeta$ Oph} are presented in Figure \ref{fig:abundances}.  The gas-phase abundances are inferred from the revised column density measurements from \citet[][and references therein]{Jenkins09}, while the solid-state elemental abundances are derived from the fourth-order polynomial spectral decomposition model.  For comparison, average abundances in young F and G stars \citep{Bensby05, Lodders09}, protosolar abundances \citep{Asplund09}, protosolar abundances with GCE \citep{Asplund09, Chiappini03}, and average abundances in B stars \citep{Nieva12} are also provided.  The elemental budget toward \object{$\zeta$ Oph} is summarized in Table \ref{tab:summary}, where we adopt protosolar abundances augmented by GCE \citep{Asplund09, Chiappini03} as the best estimate for interstellar abundances in the solar neighborhood.

For elemental Mg and Si, we calculate total abundances of $N$(Mg)/$N$(H) $=$ 51 $\pm$ 19~ppm and $N$(Si)/$N$(H) $=$ 35 $\pm$ 11 ppm, respectively.  These results are consistent with the adopted reference abundances, indicating that essentially all elemental Mg and Si along the line of sight are present in amorphous silicates.  In contrast, the detected abundance of $N$(Fe)/$N$(H) $=$ 10 $\pm$ 7~ppm suggests that $\sim$38 $\pm$ 8~ppm of elemental Fe resides in compounds other than amorphous silicates.  From our upper limit abundance estimate, we find that Fe$_{3}$O$_{4}$ could account for $\lesssim$37\% of the missing elemental Fe abundance.

In the case of elemental O, we find that the observed total (silicates plus gas) abundance of $N$(O)/$N$(H) = 433 $\pm$ 54~ppm is largely inconsistent with the adopted reference abundances presented in Figure \ref{fig:abundances}.  Under the assumption that the protosolar elemental O abundance with GCE ($N$(O)/$N$(H) $=$ 589~ppm) is representative of the current, local diffuse ISM, then as much as $\sim$156~ppm of elemental O is unaccounted for along the line of sight toward \object{$\zeta$ Oph}.  (Note that the same result is obtained if we instead adopt the average elemental O abundance in young F and G stars as the true interstellar standard.)  Our conclusions are in accord with those of \citet{Jenkins09}, who showed that spatial variations in the consumption of elemental O over many lines of sight cannot be accounted for by exchange with amorphous silicates alone.  Moreover, our limits on the abundances of H$_{2}$O ice, POM, and Fe$_{3}$O$_{4}$ imply that these materials can account for at most $\sim$37\% of the missing $\sim$156~ppm of elemental O, leaving $\sim$98~ppm of elemental O unaccounted for along the line of sight.  Thus, we have demonstrated that, other than absorption by amorphous silicates, no conclusive evidence for an {\it abundant} O-bearing solid exists in the \emph{Spitzer}-IRS or \emph{ISO}-SWS spectra toward \object{$\zeta$~Oph}.  Therefore, if a missing compound of elemental O is present in the diffuse ISM, it must not contribute significantly to the extinction opacity in the near- to mid-infrared wavelength region.

As discussed by \citet{Jenkins09}, hidden reservoirs of elemental O may reside on very large grains ($a$ $\gg$ 1~$\micron$) that are nearly opaque to infrared radiation.  However, as demonstrated in this study, the presence of thick H$_{2}$O ice mantles on grains with radii as large as 3.2~$\micron$ would introduce conspicuous structure within the 2.5--4.5~$\micron$ spectral region (see Figure \ref{fig:H2Oextinction}).  Moreover, we note that the observed interstellar extinction and polarization from ultraviolet to near-infrared wavelengths toward \object{$\zeta$~Oph} suggest no evidence for the presence of unusually large grains along the line of sight.  The star possesses an unexceptional ultraviolet extinction curve, with a somewhat weaker-than-average 2175~$\mathrm{\AA}$ extinction bump \citep{Fitzpatrick90}.  Furthermore, the ratio of total-to-selective extinction ($R_V = A_V/E_{B-V} \approx 2.6$) appears to be somewhat less than the diffuse-ISM mean (Section \ref{sec:silcomp}), suggesting a preponderance of relatively \emph{small} grains; although the wavelength of maximum linear polarization toward \object{$\zeta$~Oph} \citep[$\lambda_{\mathrm{max}} \approx 0.59$~$\micron$;][]{Serkowski75} is marginally greater than the diffuse-ISM median.  Thus, if thick H$_2$O mantles on large grains are present along the line of sight, they would contribute relatively little extinction at optical wavelengths, and very little reddening, and would be difficult to detect.

The abundance of interstellar O in all constituents of the diffuse ISM (both solid and gaseous) may be examined through the detailed structure of the 540~eV oxygen K-shell absorption-edge feature toward bright X-ray sources \citep[e.g.,][]{Paerels01, Takei02, Costantini05, Baumgartner06}.  However, the largest grains ($a > 1$~$\micron$) become optically thick to X-rays at energies near 540~eV, resulting in a nearly constant extinction cross section over the absorption-edge.  Detailed spectral decomposition analysis of this feature toward nine Galactic sightlines suggests that $\sim$15\%--25\% of the total elemental O abundance may be present in solids such as amorphous silicates and H$_{2}$O ice mantles \citep{Pinto13}.  Nevertheless, degeneracies between the absorption features produced by different species prevent the precise identification of the chemical composition of the O-bearing reservoirs.  Thus, spectral modeling of the oxygen K-shell absorption-edge offers another possibility of identifying significant repositories of elemental O on sub-micron-sized grains, but is unable to identify interstellar O that may be present in large grains ($a > 1$~$\micron$).  In conjunction with detailed spectral decomposition analyses of currently available infrared spectra, future spectroscopic observations with the {\it James Webb Space Telescope} may provide the best opportunity for detecting thick H$_{2}$O ice mantles on large grains, if they exist.

\acknowledgements
This work is based in part on observations made with the \emph{Spitzer Space Telescope}, which is operated by the Jet Propulsion Laboratory (JPL), California Institute of Technology, under a contract with NASA.  Support for this work was provided by the NASA Astrobiology Institute through contract NNA09DA80A.  B.~T.~D.~acknowledges support in part from the National Science Foundation through grant AST-1408723.  This work makes use of data products from the SWS Atlas, which was supported in part by NASA, and the Cornell Atlas of \emph{Spitzer} IRS Sources (CASSIS), a product of the Infrared Science Center at Cornell University, also supported by NASA and JPL.  The authors thank Ed Jenkins and the anonymous referees for thoughtful comments that improved the manuscript.  C.~A.~P.~is grateful to Daniel Angerhausen, Dan Watson, Tom Megeath, and Amy Stutz for insightful discussions regarding parameter degeneracy analysis.

{}

\end{document}